# On-demand manipulation of superbunching emission from colloidal quantum dots and its application in noise-resistance correlated biphoton imaging

Yunrui Song[1,2,†], Chengbing Qin[1,2,*,†], Yuanyuan Li[1,2], Xiangdong Li[1,2], Xuedong Zhang[1,2], Aoni Wei[1,2], Zhichun Yang[1,2,*], Xinghui Liu[1,2], Jianyong Hu[1,2], Ruiyun Chen[1,2], Guofeng Zhang[1,2], Liantuan Xiao[1,2,3,*], and Suotang Jia[1,2]

1. State Key Laboratory of Quantum Optics and Quantum Optics Devices, Institute of Laser Spectroscopy, Shanxi University, Taiyuan, Shanxi 030006, China
2. Collaborative Innovation Center of Extreme Optics, Shanxi University, Taiyuan, Shanxi 030006, China
3. Department of Physics and Optoelectronics, Taiyuan University of Technology, Taiyuan, 030024, China

*: Corresponding authors, email:

Chengbing Qin: chbqin@sxu.edu.cn

Zhichun Yang: yangzhichun@sxu.edu.cn

Liantuan Xiao: xlt@sxu.edu.cn

†: These authors contributed equally to this work: Yunrui Song and Chengbing Qin.

**Abstract**

Superbunching effect with second-order correlations larger than 2, $g^{(2)}(0)>2$, indicating the *N*-photon bundles emission and strong correlation among photons, has a broad range of fascinating applications in quantum illumination, communication, and computation. However, the on-demand manipulation of the superbunching effect in colloidal quantum dots (QDs) under pulsed excitation, which is beneficial to integrated photonics and lab-on-a-chip quantum devices, is still challenging. Here, we disclosed the evolution of $g^{(2)}(0)$ with the parameters of colloidal QDs by Mento Carlo simulations and performed second-order correlation measurements on CdSe/ZnS core/shell QDs under both continuous wave (CW) and pulsed lasers. The photon statistics of a single colloidal QD have been substantially tailored from sub-Poissonian distribution ($g^{(2)}(0)<1$) to superbunching emission, with the maximum $g^{(2)}(0)$ reaching 69 and 20 under CW and pulsed excitation, respectively. We have achieved correlated biphoton imaging (CPI), employing the coincidence of the biexciton and bright exciton in one laser pulse, with the stray light and background noise up to 53 times stronger than PL emission of single colloidal QDs. By modulating the PL intensity, the Fourier-domain CPI with reasonably good contrast has been determined, with the stray light noise up to 107 times stronger than PL emission and 75600 times stronger than the counts of biphotons. Our noise-resistance CPI may enable laboratory-based quantum imaging to be applied to real-world applications with the highly desired suppression of strong background noise and stray light.

## 1. Introduction

Superbunching emission, with the normalized second-order correlation larger than 2, *i.e.*, $g^{(2)}(0)>2$, indicates the existence of correlated photons or a multi-photon bundle emission and also suggests extreme photon-number fluctuations stronger than those of thermal lights ($g^{(2)}(0)=2$)[1-3]. Such highly non-classical light sources are at the heart of future quantum imaging and quantum communication networks, as well as optical quantum computers and simulations[4-7]. Prominent results have been achieved by nonlinear interactions between coherent lights and optical crystals or fibers (such as parametric down-conversion)[8-10], or random modulation of the light sources to enhance their intensity and/or polarization fluctuations via rotating ground glass or spatial light modulators[11-14]. Very recently, superbunching emission from materials or structures with nanoscale[15, 16], including semiconductor quantum dots (QDs)[17, 18], superconducting circuits[19], and generic tunnel junctions[20], has attracted much attention due to their beneficial to the integration with photonic and plasmonic devices, as well as more accessible for wafer-scale fabrication. Among them, the colloidal QDs show exciting development, where the giant superbunching emission with $g^{(2)}(0)\sim30$ from an optically driven single $CsPbBr_3$ QD at cryogenic temperature under continuous wave (CW) excitation has been achieved.[21] These developments inspire us to design practical applications in quantum imaging and holography by manipulating the superbunching effects of colloidal QDs.

Although the superbunching emission from perovskite QDs has been determined and manipulated, the comprehensive investigation on the influence of parameters (such as the lifetime of dark states, the quantum yield of the radiation emission, and the transition between the dark and bright states) to the superbunching emission remains to be discriminated. Furthermore, to our knowledge, most of the QDs-based superbunching emissions were demonstrated under the excitation of CW lasers. However, the achievement of the superbunching effect and the subsequent correlated biphoton emission under pulsed laser excitation, which can offer exact time information of the correlated photons, is highly desired for practical applications[22-24].

In this work, we comprehensively investigated the superbunching emission from the colloidal QDs by Mento Carlo simulations and disclosed the evolution of the $g^{(2)}(0)$ values varied as the lifetime of the dark exciton state and the environment temperature, as well as the quantum yield of the biexciton and bright exciton. Following the guidelines from the simulations, we synthesized CdSe/ZnS QDs with a thick shell and smooth core/shell interface to suppress the non-radiative Auger recombination and improve the quantum yield of the radiation emission. Subsequently, we achieved the on-demand manipulation of the superbunching emission of the colloidal QDs under both CW and pulsed laser excitation by varying the sampling temperature, the repetition frequency, and the excitation power, as well as the shell thickness. Photon distributions of the photoluminescence (PL) emission from a single CdSe/ZnS QD can be varied from sub-Poissonian ($g^{(2)}(0)<1$) to super-Poissonian ($1<g^{(2)}(0)\leq2$) and then to superbunching distributions, with the $g^{(2)}(0)$ values reaching 69 and 20 under CW and pulsed excitation. The Mandel-Q parameter has been used to uncover the fluctuations of the photon distributions, which is consistent with the evolution of the $g^{(2)}(0)$ values. The time difference within the biphoton, forming by the cascade emission of the biexciton and bright exciton, has been analyzed by bi-exponential functions, where the two lifetime components agree with the lifetime of the bright and dark exciton. Finally, we presented the noise-resistance correlated-photon imaging via the biphoton emitted from a single colloidal QD under a strong background and stray light illumination. The results demonstrated that correlated-photon imaging can resist stray light up to 100 times stronger than PL intensity. The resistance can be improved to close to $10^5$ by modulating the intensity of biphotons and imaging the target object through the amplitude of the Fourier transformation. These findings provide a deeper insight into the superbunching effect, and advance the potential applications through the on-demand manipulation of the superbunching emission from the colloidal QDs.

## 2. Results and discussion

2.1 *Simulation of the dynamic evolutions of the superbunching emission*

To provide deeper insight into the dynamic evolution of the superbunching emission and also guide the preparation of colloidal QDs with improved $g^{(2)}(0)$, we first disclose the behaviors of the superbunching effect under different conditions through Monte Carlo simulations. The energy-level diagram of a colloidal QD is presented in Fig. 1a. The electron in the ground state ($|G\rangle$) can be excited to excitonic states by both CW and pulsed laser. Here, two excitonic states $|BX\rangle$ and $|DX\rangle$ represent the lowest-energy bright exciton (BX, with a short lifetime) and dark exciton (DX, with a long lifetime) states. The energy difference of the bright-dark splitting ($\Delta E$) is generally in several meV. Hence, the two states can mix together, induced by thermal phonons under room temperature (with energy of about 25 meV at 298 K). However, under cryogenic conditions (such as 4.3 K with a thermal energy of about 0.37 meV), the electron populated in the dark state cannot recover to the bright state. In this case, the electron might be excited to the biexciton state ($|XX\rangle$) under the subsequent excitations, due to the long lifetime of the dark state. In this case, a cascade two-photon emission (so-called biphoton) can be demonstrated, originating from the recombination of $|XX\rangle \rightarrow |BX\rangle$ (known as biexciton, XX) and $|BX\rangle \rightarrow |G\rangle$ (known as bright exciton), as shown in Fig. 1b. The formation of biphoton and the improved ratio between biexciton and exciton result in the superbunching effect and giant $g^{(2)}(0)$, as the reports in the previous works under cryogenic conditions[17, 21]. Intuitively, the lower the temperature, the higher the ratio, and also the larger the $g^{(2)}(0)$ value. To quantitatively explore the influence of parameters (including temperatures, the lifetime of dark states, quantum yields, and so on) on the superbunching effect, we offer the Monte Carlo simulations of $g^{(2)}(0)$ under the excitation of pulsed laser (see S1 in the electronic supplementary information(ESI) for details).

Fig. 1c presents the evolution of $g^{(2)}(0)$ varied as the lifetime of the dark exciton state. As expected, the longer the dark exciton lifetime, the larger the $g^{(2)}(0)$ value (also see Fig. 1d). The colloidal QDs with longer dark exciton lifetime can be achieved by either synthesizing the high-quality crystals or varying the temperature. As reported in the previous works, as well as the results in this experiment (discussed later)[17, 21], the lower the temperature, the longer the lifetime. To disclose the influence of the temperature, we define the transition probability from the bright to dark exciton state, ξ, which can be expressed as (see details in ESI):

$$\xi = 1 - \frac{1}{e^{\Delta E/k_B T}} \quad (1)$$

where $\Delta E$ is the energy difference between the bright and dark exciton states, $k_B$ is the Boltzmann constant, and $T$ is the environmental temperature. When $\Delta E$ is fixed, the lower the temperature, the larger the ξ, and the larger the $g^{(2)}(0)$ value, as shown in Fig. 1d. Furthermore, under the same temperature (the same ξ), the longer the dark state, the larger the $g^{(2)}(0)$. We also disclose the evolutions of the $g^{(2)}(0)$ value as the quantum yields of the biexciton and bright exciton, as shown in Figs. 1e and 1f. As we expected, the larger the quantum yields, the larger the $g^{(2)}(0)$.

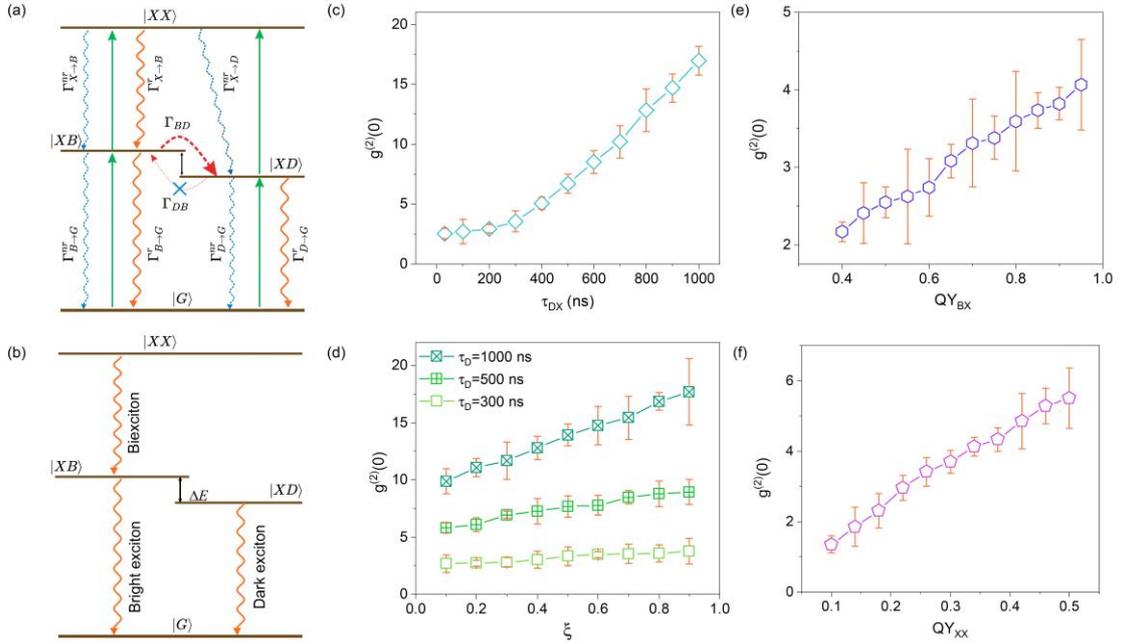

**Fig. 1 Evolutions of g²(0) via Mento Carlo simulations.** (a) Energy diagram of the colloidal quantum dots (QDs) and the possible transitions. Detailed exploration can be found in ESI. (b) Illustration of the formation of biexciton, bright exciton, and dark exciton. Evolution of g²(0) as functions of the lifetime of the dark state, $\tau_D$ (c), the transition probability, ξ (d), the quantum yields of the biexciton (e) and bright exciton (f), respectively. The errors originate from the multiple simulations.

2.2 *Design and characterization of the colloidal quantum dots*

The simulations reveal that the quantum yields and the transition between the bright and dark states play crucial roles in the improvement of the superbunching emission and the generation of correlated biphotons. Auger recombination and surface traps are the two predominant non-radiative decay pathways that will reduce the quantum yields of both biexciton and exciton emissions[25, 26]. Auger recombination will consume the energy of a photon-excited electron-hole pair (*i.e.*, exciton) to promote another carrier to a high-energy level, though the generation of negatively or positively charged exciton (all known as trion), thus reducing the formation of exciton and biexciton. On the other hand, the wave function overlapping between excitons and surface traps leads to the generation of non-radiative surface trap complexation[27, 28], which will also reduce the cascade emission of biphotons. According to our recent works, we improved the quantum yields of radiative processes through two approaches[29]. Firstly, we synthesized the gradient-alloyed CdSe/ZnS QDs with a smooth core/shell interface by lifting the shell growth temperature (from 220°C to 280°C, as shown in Fig. 2a)[30], considering that ions can diffuse into each other at higher shell growth temperature. These alloy layers result in a soft confinement potential and thus increase the biexciton quantum yield by suppressing the non-radiative biexciton Auger recombination. Secondly, we increased the thickness of the shell layer while ensuring that ZnS does not self-nucleate, to reduce the wave function overlap between the surface traps and biexciton/excitons[31, 32]. To illustrate the effectiveness of our scheme, we synthesized the core-shell layer gradient alloyed CdSe/ZnS QDs with shell thicknesses of 3 nm, 2 nm, and 1 nm, respectively, where the core size is about 10.2 nm, as the characterization of transmission electron microscopy (TEM) shown in Fig. 2b. The elemental distribution map (Fig. S2) manifests the presence of Zn elements in the core and the possession of Cd elements in the shell, indicating that the core-shell gradient alloying was achieved. We further improved the quantum yields of these QDs by applying a ligand-exchange process to replace the pristine ligand oleic acid with 1-octanethiol, and demonstrated the quantum yield to be 84% at room temperature (see Method for details).

Figure S3 presents the normalized absorption and PL spectra of the synthesized QDs with different shell thicknesses. The difference between these two spectra among the three samples can be ignored, where the PL emission centered at 642 nm at room temperature. The remaining unchanged absorption and PL of the synthesized QDs offer an excellent platform to investigate the influence of shell thickness on their superbunching effect. Fig. 2c illustrates typical PL trajectories of three single QDs with different shell thicknesses. For QDs with a shell thickness of 1 nm, the non-radiative states frequently occur, and the intensities of the bright states are relatively weak (about 13.5 thousand counts per second, kcps). In contrast, for that of 3 nm, the weak dark states appear occasionally, and the bright states hold strong intensities (27.5 kcps), indicating their higher quantum yields. PL spectra under cryogenic conditions are also displayed to explore the effect of our synthesis scheme, as shown in Fig. 2d. For QDs with a shell thickness of 1 nm, the trion peak (T) is the main component, which is much stronger than exciton (the spectral weight of T, XX, and X are 80.0%, 12.7%, and 7.3%, calculating by their integral spectral intensities. Here PL spectra of exciton comprises both bright and dark excitons). While for QDs with a shell thickness of 3 nm, the exciton emission predominates PL spectra, the trion peak almost disappears (the spectral weight of T, XX, and X are 4.4%, 11.4%, and 84.2%, respectively). Furthermore, the total intensity for QDs with 3 nm is almost twofold that with 1 nm, in accord with their PL trajectories. These characterizations indicate that CdSe/ZnS with a shell thickness of 3 nm is favorable for the improvement of superbunching emission.

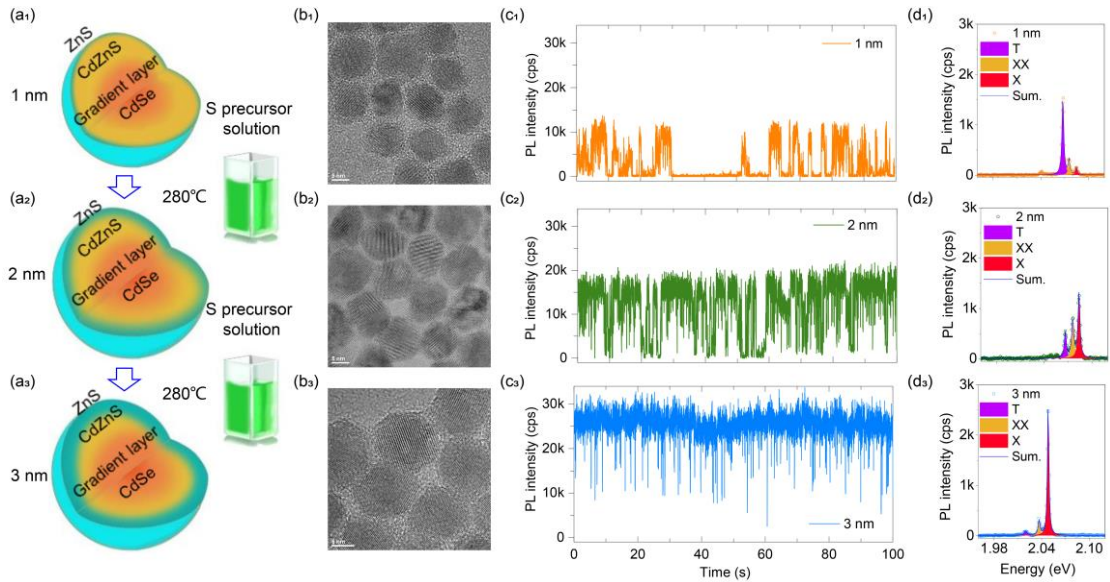

**Fig. 2 Preparation and characterization of alloyed CdSe/ZnS QDs.** (a) Schematic illustration of the synthesis of gradient alloyed CdSe/ZnS QDs with different shell thicknesses at the growth temperature of 280°C. (b) Transmission electron microscopy (TEM) micrographs for QDs with various thicknesses. Scale bar: 5 nm. (c) Typical PL trajectories of single QDs with three shell thicknesses. The excitation light was a 532 nm CW laser. (d) PL spectra of single QDs with three shell thicknesses. All the spectral profiles are deconvoluted into three peaks (Trion, T; biexciton, XX; and exciton, X) using Lorentzian functions. PL spectra of exciton comprise both bright exciton (BX) and dark exciton (DX).

2.3 *On-demand manipulation of superbunching emission under CW laser excitation*

After preparation of CdSe/ZnS QDs with different shell thicknesses, we performed second-order correlation measurements under CW laser excitation with varying temperatures, as shown in Fig. 3. The detailed experimental setup for the optical system can be found in Method. Briefly, CdSe/ZnS QDs spin-coated on $SiO_2$/Si substrates were loaded in a liquid-helium-free cryogenic vacuum chamber, where the temperature can be varied from 4.3 K to room temperature (298 K). The excitation light was a 532 nm CW laser. PL emission from single QDs was collected by a high numerical aperture cryogenic objective and split equally into two single-photon detectors (SPDs) to conduct Hanbury Brown and Twiss (HBT) measurements[33, 34]. Fig. 3a presents $g^{(2)}(\tau)$ of a single QD with a shell thickness of 3 nm under different temperatures. Under room temperature, an anti-bunching phenomenon with $g^{(2)}(0)$ about 0.03 can be demonstrated, indicating its single particle feature. With the decrease of the temperature, the anti-bunching phenomenon disappears, and the value of $g^{(2)}(0)$ gradually increases and tends to 1. For example, under 30 K, $g^{(2)}(0)$ closes to 1, and PL emission behaviors coherent-like emission. Further cooling results in the occurrence of bunching effects (*e.g.*, $g^{(2)}(0)$=1.5 at 20 K). Superbunching emission with $g^{(2)}(0)$>2 generally emerges with a temperature lower than 18 K, where a shape peak can be demonstrated at the time delay between two SPDs equaling to zero, indicating strong bundle emission of biphotons and short lifetimes of the PL spectra. The maximum $g^{(2)}(0)$ observed in this work reaches 69, as illustrated in Fig. 3b. To our knowledge, it is also the maximum value for colloidal QDs with superbunching emission. The evolution of $g^{(2)}(0)$ varied as the inverse of the temperature can be reasonably fitted by mono-exponential function, as the solid line shown in Fig. 3c. On the other hand, with the increase of the excitation power, $g^{(2)}(0)$ decreases (Figs. 3d and S4) and the evolution behaviors are consistent with the simulation and previous works. These phenomena demonstrate that the photon emission in the colloidal QDs can be substantially tailored from sub-Poissonian to super-Poissonian and superbunching distributions.[35, 36]

To confirm the universality of the superbunching emission in the CdSe/ZnS QDs, we performed the second-order correlation measurement on different single QDs and recorded their maximum $g^{(2)}(0)$. Figures 3e-3g present the histograms of $g^{(2)}(0)$ for QDs with shell thicknesses of 1 nm, 2 nm, and 3 nm, respectively. The averaged $g^{(2)}(0)$ values of these three types of colloidal QDs have been demonstrated to be 6.6, 19.4, and 33.9, respectively, through Gauss fits of the histograms (as the solid lines shown in the figures). The increased tendency of the $g^{(2)}(0)$ values can be attributed to the improved quantum yields and cascade biphoton emission for the thicker shell alloyed QDs, consistent with our expectation and Mento Carlo simulations. These results prove that the superbunching emission can be manipulated on-demand by improving the quality of the colloidal QDs, the environment temperature, and the excitation powers.

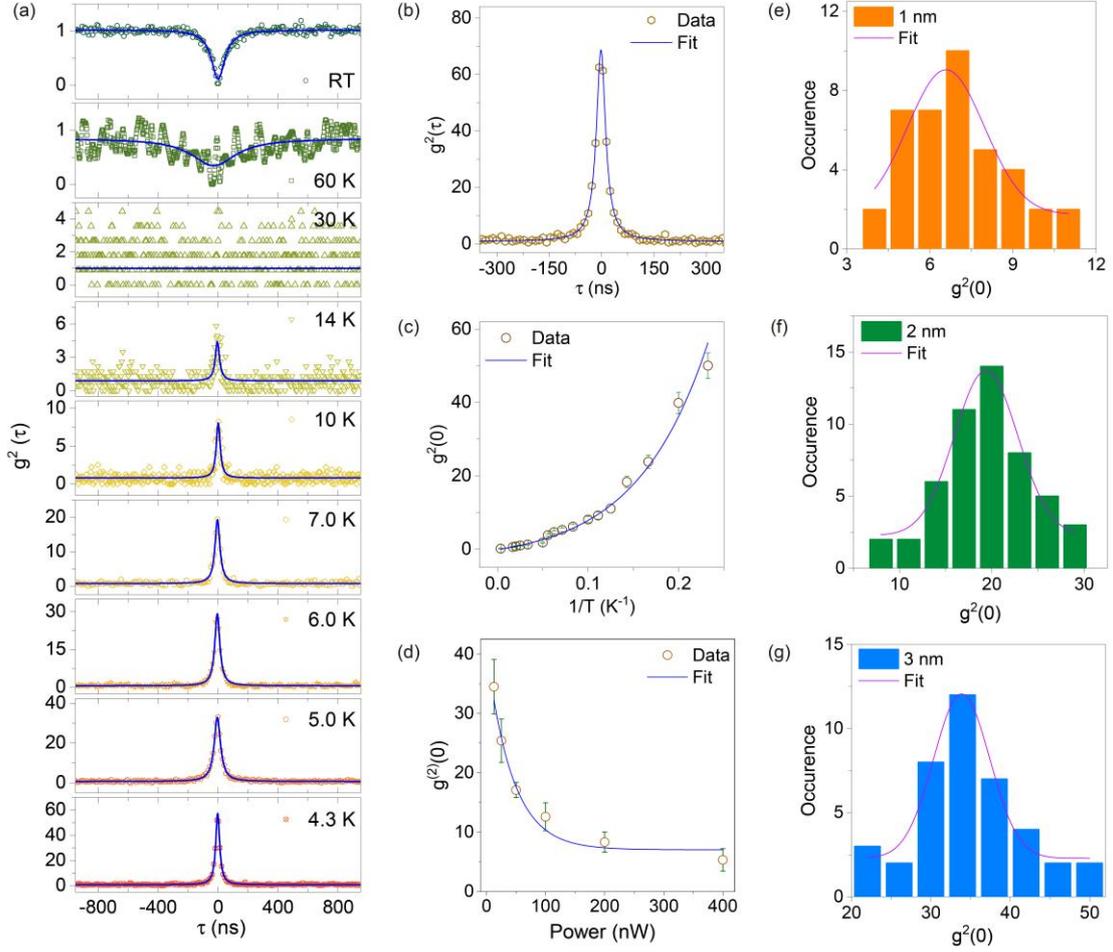

**Fig. 3 Manipulation of the superbunching emission of colloidal QDs under CW laser excitation.** (a) $g^{(2)}(\tau)$ for a single CdSe/ZnS QD under various temperatures. The excitation light was a 532 nm CW laser with a power of 3.8 nW. (b) The maximum $g^{(2)}(0)$ up to 69 observed in the experiment. The solid lines in a and b are fit results. (c) The $g^{(2)}(0)$ varied as the inverse of temperature (1/T). The solid line is a mono-exponential fit, $g^{(2)}(0) = a \times e^{\alpha/T} + b$, with the exponential factor α being 12.7 K and the correlation coefficient ($R^2$) being 0.964. (d) The $g^{(2)}(0)$ as a function of the excitation power. Histograms of the maximum $g^{(2)}(0)$ of single QDs with the thickness of 1 nm (e), 2 nm (f), and 3 nm (g), respectively. The solid lines are the Gauss fits.

2.4 *On-demand manipulation of superbunching emission under pulsed laser excitation*

Manipulation of the superbunching emission under pulsed laser excitation is deeply desired, due to that the arriving time of each emission photon, as well as the time difference between the biexciton and bright exciton (Δ*t*), can be clearly determined, which is highly beneficial for practical applications, such as correlated-photon imaging (also known as heralded-photon imaging)[37], and communications[38]. However, it is rather challenging to address a big $g^{(2)}(0)$ under pulsed excitation because of its strong peak power density (similar to Fig. 3d). In our scheme, the high-quality QDs allow us to manipulate the superbunching effect under pulsed excitation, as shown in Fig. 4. Similar to the CW excitation, the anti-bunching phenomenon can be determined at room temperature for a single QDs under pulsed excitation (Fig. 4a). With the decrease of the temperature, the curves of $g^{(2)}(\tau)$ convert from the anti-bunching to thermal radiation to superbunching emission (at about 14 K). In view of the limited photons detected in each measurement, the coincidence counts that occur at time delay τ=±*NT* (with *N* as an integer and *T* as the period of the pulsed laser) are insignificant. To explore the side coincidences distinctly, we increased the integration times and enlarged the $g^{(2)}(\tau)$ curves under different repetition frequencies, as shown in Fig. 4c. Compared with the CW excitation, although the maximum $g^{(2)}(0)$ value under pulsed excitation presents a slight decline, the $g^{(2)}(0)$ varied as the inverse of temperature also follows mono-exponential function, as the solid line shown in Fig. 4b, definitely consistent with the CW excitation (Fig. 3c). We also note that the maximum $g^{(2)}(0)$ values for a single QDs almost linearly increase with the repetition frequency of the pulsed laser, as shown in Fig. 4d, hinting that the largest $g^{(2)}(0)$ should be demonstrated under CW excitation.

Superbunching emission under pulsed excitation also allows us to disclose the relationship between $g^{(2)}(0)$ and the lifetimes of PL emission as well as counts of biphotons. Fig. 4e offers the time-resolved PL spectra (TRPS) of single QDs under different temperatures. The curves can be fitted by bi-exponential or tri-exponential functions (Fig. S5), where the shortest lifetime component can be attributed to the biexciton ($\tau_{XX}$~0.5 ns), the middle one to the bright exciton ($\tau_{XB}$~3 ns), and the longest one to the dark exciton ($\tau_{XD}$~100 ns). The corresponding lifetimes and spectral weights (dividing the integral intensity of each component by the total intensity) have been presented in Fig. 4f and Table S1. Generally, the weight of the biexciton is pretty small, and its lifetime is too short (close to the instrument response function). Thus, only the components of biexciton under 4.3 K and 5.0 K have been demonstrated. Note that the TRPS curve under 4.3 K manifests a predominate short lifetime component, indicating a massive formation of the biexciton and bright exciton that is highly beneficial to the generation of cascade emission and thus the superbunching effect. We also counted the number of biphotons (*i.e.*, both SPDs give responses in one laser pulse) and statistics their time differences ($\Delta t$), as shown in Fig. 4e and S6, respectively. With the increase of the temperature, the evolution of the counts for biphotons varying as the temperature agrees well with that of $g^{(2)}(0)$, as presented in Fig. 4g, suggesting the direct connection between the superbunching effect and the number of biphotons. Furthermore, the histograms of the time difference follow the bi-exponential functions. The fitting results are shown in Fig. S6 and Table S2. The lifetimes of the two components ($\tau_1$ and $\tau_2$) agree reasonably well with the lifetimes of bright and dark excitons, as shown in Fig. 4f, proving that the cascade emissions of both biexciton-bright exciton and biexciton-dark exciton contribute to the formation of biphotons.

To further reveal fluctuation features of the superbunching emission in colloidal QDs, we evaluated the Mandel-Q parameter ($Q(t)$) of PL emission, which is a crucial parameter to characterize the photon number fluctuation in the quantum systems and can provide information about coherence and correlation with profound physical implications[39]. The formula of $Q(t)$ can be expressed as[40]:

$$Q(t) = \frac{\left\langle \left(\Delta N(t)\right)^2 \right\rangle - 1}{\left\langle N(t) \right\rangle} \quad (2)$$

Here $N(t)$ is the total photon number in the period $t$, which can be obtained by counting the emitted photons from the time-correlated single-photon counting (TCSPC) data acquisition card; the angle bracket $\langle N(t) \rangle$ can be understood as a mean photon number in the same period *t*, and $\langle (\Delta N(t))^2 \rangle$ is the variance of *N(t)*. For *Q(t)*=0, the light field shows a Poissonian distribution, meaning that the light field manifests a coherent emission behavior. While *Q(t)*>0 and *Q(t)*<0 represent a super- and sub-Poissonian distribution, respectively[41, 42]. Fig. 4h presents the calculated temperature-dependent *Q(t)* of a single QD at different time scales, demonstrating a completely different behavior from that of the two-level system, such as single molecules (Fig. S7). On the time scale of the initial photon detection (within 3.5 ns for the lifetime of the biexciton and bright exciton), the light field of PL emission at 4.3 K presents a super-Poissonian distribution (*Q(t)*=6.7×10$^{-4}$), implying a large number of associated photon pairs and strong photon correlation. With the increase of the temperature, *Q(t)* gradually decreases and tends to 0. For example, *Q(t)*=-1.2×10$^{-5}$ at 18 K, indicating a weak single-photon emission feature rather than a strong cascade biphoton emission. On the other hand, with the increase of the time scale, *Q(t)* presents a declining tendency, practically for *t*~100 ns, *Q(t)*<0 for most cases. This phenomenon can be attributed to the contribution of the dark exciton due to the fact that the lifetime of the dark exciton is in the time scale of 100 ns (Fig. 4f). Thus, at this time region, the predominate contributions to *Q(t)* varies from the correlated biphotons to dark exciton emission, which behavior a single-photon source and manifests a sub-Poissonian distribution. After this inflection point, *Q(t)* continues to rise, and such strong fluctuations arise not only from the dark exciton emission but also from the signal from arbitrary photons beyond the dead time of SPD (such as photons from different pulses). Overall, the evolution of *Q(t)* as a function of temperature and time scale offers more profound insight into the mechanism of PL emission from sub-Poissonian distribution to the super-bunching effect and also provides a fertile ground for future applications.

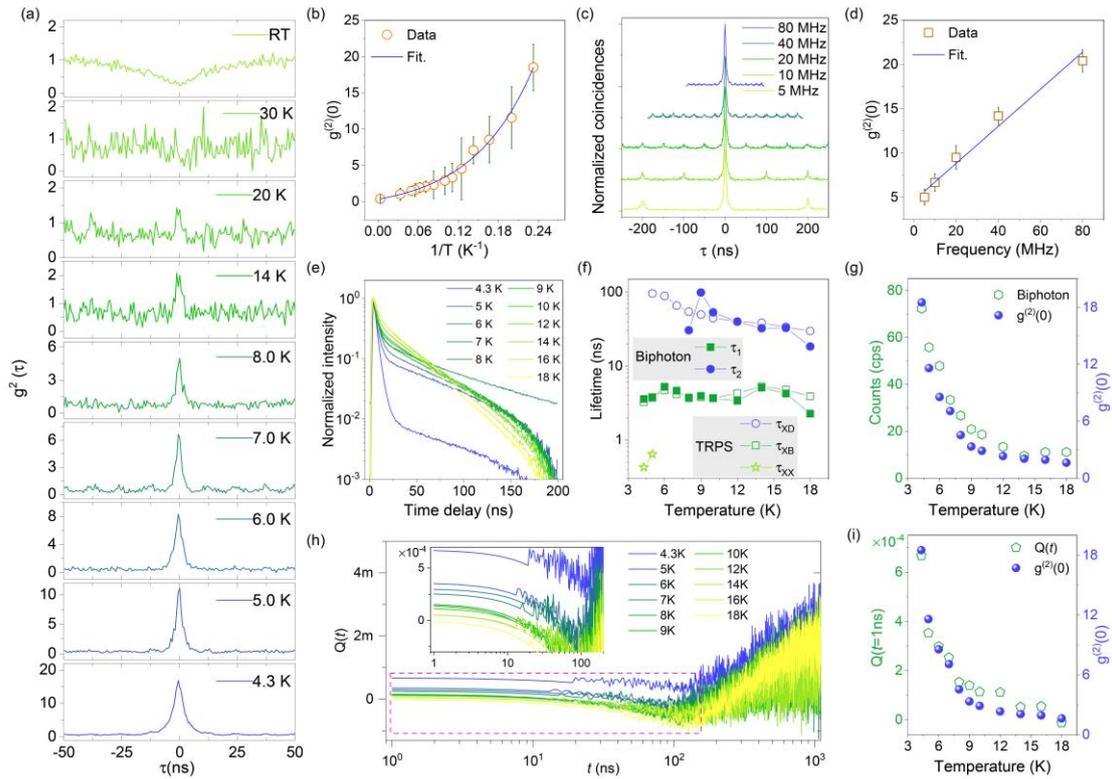

**Fig. 4 Manipulation of the superbunching emission of colloidal QDs under pulsed laser excitation.** (a) $g^{(2)}(\tau)$ for a single CdSe/ZnS QD under various temperatures. The excitation light was a 532 nm pulsed laser. (b) The $g^{(2)}(0)$ varied as the inverse of temperature (1/T). The solid line is the mono-exponential fit, $g^{(2)}(0) = a \times e^{\alpha/T} + b$, with the exponential factor α being 11.0 K and the correlation coefficient ($R^2$) being 0.992. (c) $g^{(2)}(\tau)$ for a single CdSe/ZnS QD under the excitation of various repetition frequencies at 4.3 K with a similar excitation power. (d) $g^{(2)}(0)$ varied as repetition frequencies. The solid line is a linear fit. (e) Time-resolved PL spectra (TSPR) of a single CdSe/ZnS QD under different temperatures. (f) Comparisons of lifetime components between TSPR and time delays of biphotons. $\tau_{XX}$, $\tau_{XB}$, and $\tau_{XD}$ represent the lifetimes of biexciton, bright exciton, and dark exciton, respectively. (g) Comparisons between counts of biphoton per second and $g^{(2)}(0)$ values under different temperatures. (h) Temperature-dependent Mandel-Q parameters (Q(t)) at various time scales. The inset shows the enlargement of the dashed region. (i) Comparisons between $Q(t=1\ ns)$ and $g^{(2)}(0)$ values under different temperatures.

2.5 *Noise-resistance correlated biphoton imaging based on the superbunching emission*

Correlated photon sources or entangled photon pairs offer great superiority in quantum illumination, which can effectively suppress background and stray lights[43-45]. Previous quantum illuminations were mainly achieved via spontaneous parametric down-conversion (SPDC) protocols with bulk crystals[23, 24, 46]. To show the potential applications of the superbunching effect and the advantage of the colloidal QDs with the lateral scale of nanometers, we performed correlated biphoton imaging in the presence of strong uncorrelated background noise. The experimental setup is presented in Fig. 5a. The colloidal QDs loaded in the cryogenic chamber were excited by a pulsed laser. The biphoton emission from the chamber was split into two arms by a beam splitter (BS, 50:50). For the reflecting arm (also named the signal arm), the photons focused on a spatial light modulator (SLM), which was used to simulate target objects. In this experiment, three letters "SXU" and a cross mark "+" with high transmittance were loaded on the SLM, while the transmittance of the other regions was set to zero. The photons were focused and collected by a 4*f* system through two objectives with the same NA (NA=0.6, for Obj1 and Obj2) and then detected by an SPD (SPD1). A background light field was deliberately introduced using a thermal light source to illuminate the SPD1 directly, which overlapped the reflection photons by a dichroic mirror (DM). The thermal light source can simulate real-world environmental noise and broad-band illumination. To improve the fluctuations of environmental noise, we modulated the thermal light source by a random noise signal. For the transmitted arm (also named the reference arm), two objectives (Obj3 and Obj4) were used to form another 4*f* system to construct an identical optical path. The transmitted photons passed through the 4*f* system were then detected by another SPD (SPD2). The relative arrival time delay between the two paths was eliminated by a variable optical delay line (M3, HR, and M4).

Without noise, the target object can be determined by raster scanning SLM and recording the photon counts of SPD1, which is also named single-photon imaging (SPI).

With the illumination of the thermal light, the correlated biphoton imaging (CPI) was determined by counting the biphoton events, which is identified as that when both SPD1 and SPD2 responded in the same pulse period and the time difference between the two electrical responses within Δ$t$. Here, Δ$t$ is set as the lifetime of the bright exciton, *i.e.*, Δ$t$=3.5 ns at 4.3 K. Fig. 5b presents the distributions of the time difference of biphotons for both colloidal QDs and the thermal noise. Note that the distribution of colloidal QDs is close to zero and manifests a short and a long lifetime, which has been discussed in Fig. 4f. While for the thermal noise, the time difference distributes within the entire pulse period and is very sparse due to their random nature. To take advantage of the noise-resistance feature of the correlated biphoton, three letters "SXU" under intense stray light were imaged by raster scanning SLM, where the ratios between the counts of noise and the PL emission of a single colloidal QD ($N_{\text{Noise}}/N_{\text{PL}}$) were varied from 2.45:1 to 104:1, the results are shown in Figs. 5c-5d and Table S3, respectively. Apparently, for conventional SPI, the information (SXU) is overwhelmed by strong stray light (Fig. 5c$_3$). Nevertheless, CPI can eliminate the stray light by extracting the correlated intensity (Fig. 5c$_4$). In fact, for a single colloidal QD under the excitation of a 40 MHz pulsed laser at 4.3 K, the averaged counts of biphoton within 100 ms (counts per pixel) can reach close to 10 with $g^{(2)}(0)$ up to 8.5. The counts of PL of the colloidal QDs can reach about 500 (Table S3). In this case, CPI can effectively suppress stray light that is close to 100 times stronger than its PL intensity and more than 6000 times stronger than the counts of biphoton. To quantitatively compare the results, we calculated the contrast-to-noise ratios (CNR) for all the images[46], which can be expressed as:

$$CNR = \frac{|\bar{I}_n - \bar{I}_s|}{\sqrt{\sigma_n^2 + \sigma_s^2}} \quad (3)$$

where $I_s$ and $I_n$ are the intensities of the signal area (*i.e.*, target objects) and noise (*i.e.*, the background and stray light), $\bar{I}_n$ and $\bar{I}_s$ are the corresponding mean values, $\sigma_n$ and $\sigma_s$ are the corresponding standard deviations. Fig. 5c$_5$ illustrates the calculated CNR for SPI and CPI, where the values for SPI decrease rapidly from 2.39 to 0.02, with the $N_{\text{noise}}/N_{\text{PL}}$ increasing from 2.45 to 104, while for CPI, these values only decrease from 2.12 to 0.63. showcasing the superiority of CPI. We also performed CPI for the same single QD under different temperatures, as shown in Fig. 5e and Table S3. As expected, with the increases in the temperature, the quality of CPI degrades, due to that the counts of the biphoton decrease. Even though, a reasonably distinct picture can still be achieved at 7.0 K, with the stray light 20 times stronger than the PL intensity of the colloidal QDs (Fig. 5e$_3$). To improve the imaging quality further, we performed intensity modulation to the QDs' PL emission through an acousto-optic modulator (AOM, as shown in Fig. 5a). In this case, the arriving times of PL (and also biphotons) can be modulated by a sinusoidal wave. After performing discrete Fourier transformation (DFT) to the arriving time of each photon, a significant signal can be found in their frequency spectra, as shown in Fig. 5f. While for stray light, no distinguished signal can be found due to their white noise feature. Hence, we can detect the target object by combining the time-correlated biphoton and intensity modulation and perform the imaging via the amplitude in the frequency domain of the correlated biphoton, named frequency-domain CPI (FD-CPI). The results are presented in Fig. 5g. The target object can still be visualized with the colloidal QDs under different temperatures, even though the intensity of the stray light is almost two orders of magnitude stronger than their PL emission. For example, under $N_{\text{noise}}/N_{\text{PL}}$~106 and the temperature of 18 K, FD-CPI still can present good contrast, as shown in Fig. 5g$_3$. We also calculated the counts of biphotons, the intensities of the stray light, PL emission, and their amplitudes, as well as the corresponding CNR, which are listed in Tables S4 and S5. From the calculated values and Fig. 5h, we can conclude that a reasonably good picture can be determined via FD-CPI even though the background and stray light noise are 75600 times stronger than that of biphotons, indicating the strong noise-resistance feature of FD-CPI.

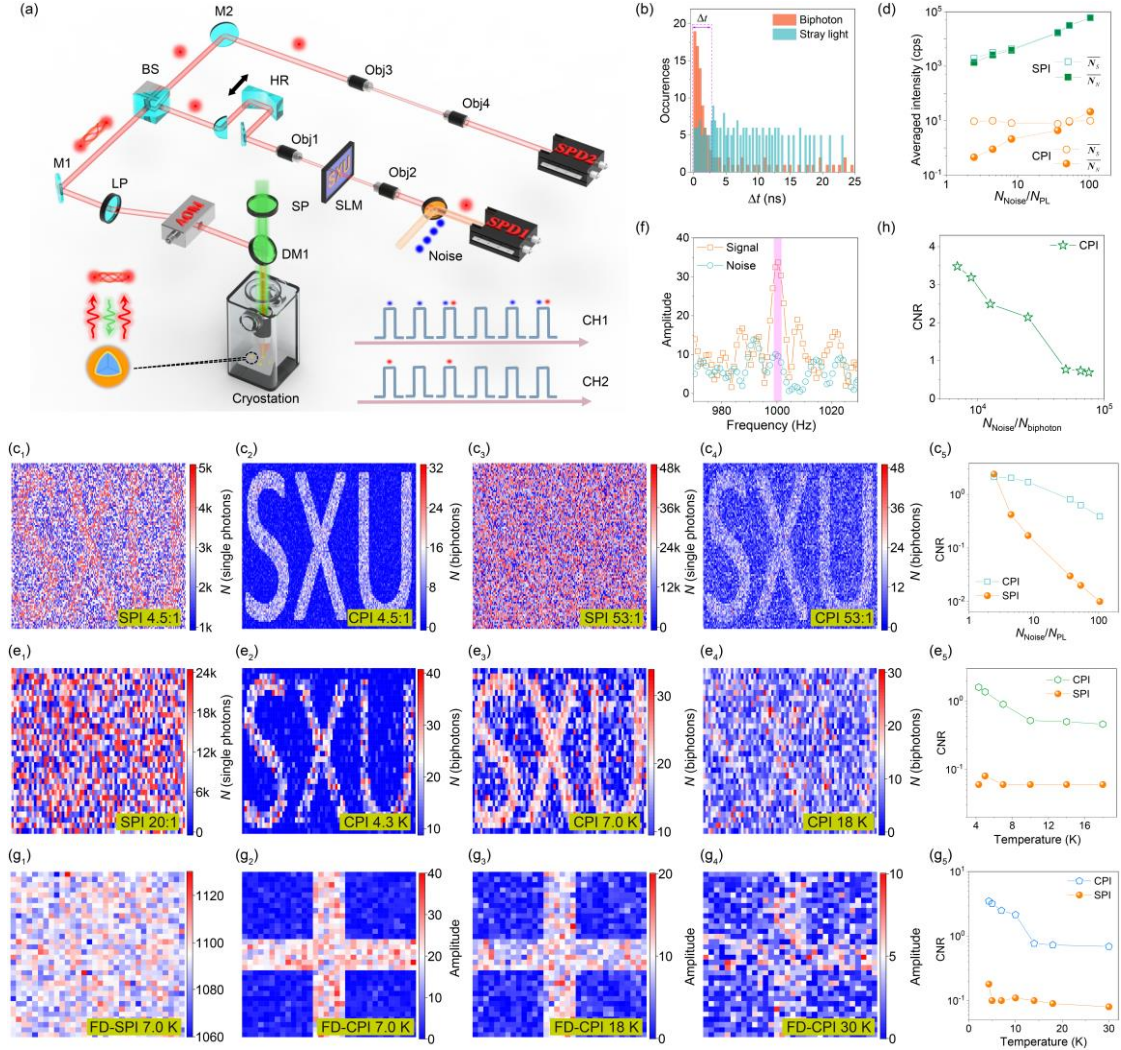

**Fig. 5 Noise-resistance features of the correlated biphoton imaging (CPI)**. (a) Schematic illustration of the experimental setup. SP/LP: short/long pass filter, DM: dichroic mirror; AOM: acousto-optic modulator; M: mirror; BS: beam splitter; HR: high reflection mirror; Obj: objective; SLM: spatial light modulator; SPD: single photon detector. (b) Statistics of the time difference for the two electric responses in one laser pulse. The dashed rectangle represents the count of biphoton with a time difference of 3.5 ns. The PL intensity of the colloidal QDs (for biphoton) was about 500 cps. The intensity of the background and stray light was about 50 kcps. (c) Single photon imaging (SPI) and CPI of "SXU" under $N_{\text{nosie}}/N_{\text{PL}}$ of 5:1 and 50:1, respectively, and the calculated CNR varied as $N_{\text{noise}}/N_{\text{PL}}$. The integral time for each pixel was 100 ms. Images are 91×200 pixels. (d) Averaged counts per pixel of noise ($\overline{N_n}$) and signal ($\overline{N_s}$) for SPI and CPI as a function of $N_{\text{noise}}/N_{\text{PL}}$. (e) Temperature-dependent SPI and

CPI under $N_{\text{noise}}/N_{\text{PL}}$=25:1 and the corresponding CNR. Images are 30×69 pixels. (f) Amplitude of the Fourier transformation spectra of modulated biphoton and unmodulated background, a significant signal for modulated biphoton, has been highlighted by a purple dashed area at the frequency of 1000 Hz. (g) Temperature-dependent Fourier-domain SPI (FD-SPI) and -CPI (FD-CPI) under $N_{\text{noise}}/N_{\text{PL}}$=100:1, and the corresponding CNR. The integral time for each pixel was 1000 ms. Images are 30×69 pixels. (h) CNR of FD-CPI as a function of $N_{\text{noise}}/N_{\text{PL}}$.

## 3. Conclusion

We have demonstrated dynamic evolutions of the superbunching effect in the colloidal QDs via Mento Carlo simulations, and performed temperature-dependent second-order correlation measurements on CdSe/ZnS colloidal QDs with different shell thicknesses under both CW and pulsed laser excitation. The maximum $g^{(2)}(0)$ values up to 69 and 20 for CW and pulsed excitation have been determined under 4.3 K. The evolution of $g^{(2)}(0)$ varied as temperature, excitation powers, shell thicknesses (strongly associating with quantum yields of the biexciton and bright exciton), and the repetition frequency of the pulsed laser. The results proved that the superbunching effect of the colloidal QDs can be manipulated as required. We also performed noise-resistance correlated biphoton imaging via the counts and Fourier transformation amplitude of biphoton emitted from a single CdSe/ZnS colloidal QDs. Reasonable good contrast results, unde the ratio of $N_{\text{noise}}/N_{\text{PL}}$ ($N_{\text{noise}}/N_{\text{biphoton}}$) up to 104 (6061), have been determined. By performing intensity modulation, these ratios can reach 106 (75600), indicating the excellent noise-resistance feature of correlated biphoton imaging via colloidal QDs. Through resilience to the background and stray noise, our protocol may have promising applications in real-world implementations, including quantum illumination, quantum radar, and quantum microscopy for low-light-level imaging. Improvements in the counts of biphotons, such as fabrication of QDs arrays and advancement in collection efficiency (such as microlens and microcavity), should enable quantum imaging applications to be realized and applied outside the laboratory.

**Methods:**

**Sample preparation**: Alloyed CdSe/ZnS QDs with a gradient in chemical composition were synthesized by a conventional "thermal injection" method. In a typical synthesis process, 0.4 mmol of CdO, 10 mmol of Zn(OAc)$_2$, and 15 mL of oleic acid (OA) were loaded into a 100 mL three-necked flask. After degassing at 120°C for 40 min, a mixture of Zn(OA)$_2$ and Cd(OA)$_2$-precursors was obtained. Then, 30 mL of 1-octadecadienyl (ODE) was added into the flask, and the solution was heated under an argon atmosphere to 320°C. The Se precursor (0.8 mmol Se dissolved in 0.4 mL of tri-octylphosphine (TOP)) was then rapidly injected into the flask. The solution was kept for 30 min to obtain the alloy core. The temperature was then set to 310°C, and another Se precursor solution (0.5 mmol Se dissolved in 0.25 mL TOP) was added to the flask. The solution was held for 10 min to allow the growth of the intermediate layer. Then, Cd precursor solution (1.4 mmol Cd(OA)$_2$ in 7 mL ODE) and S precursor solution (2.8 mmol S in 1.4 mL TOP) were added to the flasks simultaneously dropwise at rates of 10.5 mL h-1 and 2.1 mL h-2, respectively, at 280°C to achieve the growth of CdSeS shells. Finally, the S precursor solution (1.0 mmol S in 0.5 mL TOP) was added to the flask, and the solution was kept for 20 min to achieve ZnS shell growth.

When the desired alloyed CdSe/ZnS QDs were obtained, the solution was cooled to room temperature by an ice bath. The resulting alloyed CdSe/ZnS QDs with OA ligands were washed three times with ethanol and dispersed in heptane. For simplicity, the alloyed core and the intermediate layer are collectively referred to as the core, and the CdZnS and ZnS layers are collectively referred to as the shell. The specific steps were as follows: the purified CdSe/ZnS QDs with OA ligands were first dissolved in chlorobenzene. Then, a certain amount of 1-octanethiol was added to the solution. After continuous stirring at 50°C under an argon atmosphere for 14 h, the obtained CdSe/ZnS QDs with 1-octane thiol ligand were washed with ethanol and dispersed in heptane. Similarly, alloyed CdSe/ZnS QDs with an oleylamine ligand can be obtained by a ligand exchange reaction.

**Experimental setup for second-order correlation measurements.** All the experiments were performed on a homemade confocal scanning microscope. A continuous wave (CW) laser ($\lambda$=532 nm, CNIlaser, MGL-U-532) and a pulsed laser (Picoquant, LDH-D-FA-530L) were used as the excitation sources. The repetition frequency of the pulsed laser can be varied from 10 kHz to 80 MHz. After passing through a beam expander (Thorlabs, BE06R/M), the excitation laser was reflected by a dichroic mirror (Semrock, Di03-R532-t3-25×36) and a 4*f* telecentric relay lens system with a fast-steering scanner (Newport, FSM-300-M-03). Eventually, the laser was focused on the sample through a cryogenic objective lens (Attocube LT-APO, NA=0.82), where the sample was placed on a cryogenic scanning stage (Attocube ANC350). The temperature of the cryogenic chamber (Montana Instruments, 4200-100-Cryostat) can be continued tuned in the range of 4.3 K-350 K. PL emission was collected by the same objective lens, passed through a long pass filter (Semrock, LP03-532RE-25). Subsequently, PL photons were split evenly into two beams by a beam splitter (with a ratio of 50:50). Two beams of PL photons were detected by two single-photon detectors (SPD, PicoQuant, τ-SPAD-50). The arrival times of the photon sequence were recorded by a time-correlated single-photon counter (TCSPC, PicoQuant, HydraHarp 400). The second-order correlations were determined by counting the coincidences of the two channels with different time delays.

**Acknowledgment**

The authors gratefully acknowledge support from the National Key Research and Development Program of China (Grant No. 2022YFA1404201), Natural Science Foundation of China (Nos. U22A2091, 62222509, U23A20380, 62127817, 62075120, 62075122, 62205187, and 62105193), Shanxi Province Science and Technology Innovation Talent Team (No. 202204051001014), the key research and development project of Shanxi Province (202102030201007), and 111 projects (Grant No. D18001).

**Author contributions**

C. B. Q., Z. C. Y., and L. T. X. conceived and designed the experiment. Y. R. S., C. B. Q, Y. Y. L., and X. D. L. conducted the experiments and analyzed the data. Y. R. S., X. D. Z., A. N. W., and Z. C. Y. contributed to the optical setup and sample preparation. X. H. L., R. Y. C., and S. T. J. contributed to the electronic measurements and data analysis. J. Y. H. and G. F. Z. contributed to the theoretical analysis and simulations. C. B. Q., Z. C. Y., and L. T. X. wrote the manuscript with input from all other authors. C. B. Q. and L. T. X. supervised the project.

**Competing interests.**

The authors declare no competing interests.

**Data and materials availability.**

All data needed to evaluate the conclusions in the paper are present in the paper or the supplementary materials.

**Electronic supplementary information (ESI).**

ESI includes